\documentclass[10pt,conference]{IEEEtran}
\IEEEoverridecommandlockouts
\usepackage{cite}
\usepackage{amsmath,amssymb,amsfonts}
\usepackage{algorithmic}
\usepackage{graphicx}
\usepackage{textcomp}
\usepackage[table,xcdraw]{xcolor}
\usepackage{multirow}
\usepackage{caption}
\usepackage{subcaption}
\DeclareGraphicsExtensions{.png}
\DeclareGraphicsExtensions{.jpg}
\graphicspath{ {./figs/} }

\def\BibTeX{{\rm B\kern-.05em{\sc i\kern-.025em b}\kern-.08em
    T\kern-.1667em\lower.7ex\hbox{E}\kern-.125emX}}
\begin{document}

\title{\huge{Encoder-Quantization-Motion-based Video Quality Metrics}}

\author{Yixu Chen, Zaixi Shang, Hai Wei, Yongjun Wu, Sriram Sethuraman\\
\textit{Amazon Prime Video}
}

\maketitle

\begin{abstract}
In an adaptive bitrate streaming application, the efficiency of video compression and the encoded video quality depend on both the video codec and the quality metric used to perform encoding optimization. 
The development of such a quality metric need large scale subjective datasets. In this work we merge several datasets into one to support the creation of a metric tailored for video compression and scaling. We proposed a set of HEVC lightweight features to boost performance of the metrics. Our metrics can be computed from tightly coupled encoding process with 4\% compute overhead or from the decoding process in real-time. The proposed method can achieve better correlation than VMAF and P.1204.3. It can extrapolate to different dynamic ranges, and is suitable for real-time video quality metrics delivery in the bitstream. The performance is verified by in-distribution and cross-dataset tests. This work paves the way for adaptive client-side heuristics, real-time segment optimization, dynamic bitrate capping, and quality-dependent post-processing neural network switching, etc. 
\end{abstract}


\section{Introduction}

While subjective video quality, captured through Mean Opinion Scores (MOS) by Absolute Category Rating (ACR) study, offer a precise method for gauging human perceptions of video quality, but they come with their own challenges. These assessments are often labor-intensive, costly, non-scalable. For the purpose of segment-level video encoding optimization, quality of service monitoring, online streaming providers are in search of objective Video Quality Metrics (VQM) that align closely with subjective ratings while keep the real-time run time requirement for low latentcy application. 

\section{Related Work}
\label{sec:lit_survey}
Compared to traditional metrics like PSNR, SSIM\cite{SSIM, MSSIM, NIQE}, VMAF\cite{li2016vmaf} has shown superior correlation and became a preferred choice in the industry. However, the runtime efficiency of pixel-based VMAF is much slower than the high frame rate real-time requirement for latency sensitive applications. 

Among the VQM models from ITU-P.1204 standard, the most lightweight one is P.1204.3 \cite{p1204} bitstream-based metric which uses features from the decoding process for quality prediction, e.g. quantization, motion, frame rate, bitrate. It can be computed from the decoding process of the video bitstream. Compared to pixel-based features, bitstream features are more agnostic to color gamut and transfer function difference between SDR and HDR contents. Our benchmark shows P.1204.3 performs better on HDR video datasets than pixel-based VMAF when they both trained on SDR dataset and vice versa. 

\begin{figure}[h]
\centering
\includegraphics[width=0.7\linewidth]{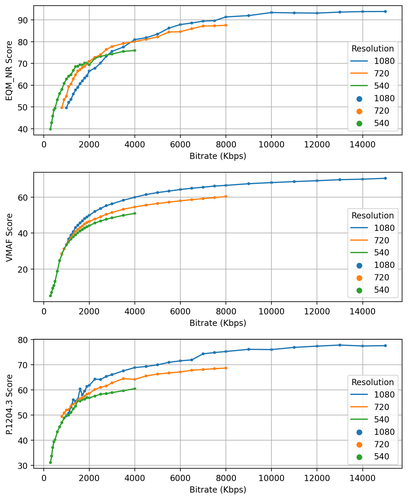}
\caption{R-Q curve of proposed EQM, P.1204.3 and VMAF on a 1-minute sports video. EQM could accurately capture the crossover among 3 resolutions.}
\label{fig:RQs}
\end{figure}

However both VMAF and P.1204.3 struggle to accurately predict the resolution bitrate crossovers and often prefer higher resolution videos from Figure \ref{fig:RQs}. This is due to P.1204.3 relies on a simple parametric function to model quality degradation from scaling, which should be content dependent. 

In additional, P.1204.3 can be slower than VMAF due to the single-threaded decoding requirement. 
To accommodate the low latency requirement, we introduce an encoder-side video quality metric with a redesigned feature set to better capture contrast masking, motion masking, MV-derived saliency-weighted quantization. The proposed method is trained on a large-scale heterogeneous video quality dataset.
The EQM model enables applications like adaptive bitrate streaming heuristics, quality-dependent neural network switching, enhanced quality monitoring, dynamic bitrate capping, etc.

\section{HDR Combined Dataset}
\label{sec:datasets}
We have created 4 HDR datasets and 1 SDR dataset in collaborations with the University of Texas Austin Laboratory for Image and Video Engineering (LIVE). The 4 HDR datasets are: HDR-LIVE\cite{hdr_live}, HDR-AQ\cite{hdr_aq}, HDR-Sports\cite{hdr_sports}, and HDR-SDR\cite{hdr_sdr}. The SDR dataset is Low-Bitrate Sports (LBS)\cite{lbs}. HDR-LIVE\cite{hdr_live} is to study the ambient illumination's impact on video quality. HDR-AQ\cite{hdr_aq} is to study the relation between adaptive quantization (AQ) and video quality. HDR-Sports\cite{hdr_sports} is to include different frame rate variations and anchor videos to combine previous datasets. HDR-SDR\cite{hdr_sdr} is to compare HDR vs SDR versions of the same videos and study the quality impact for different TV panels. LBS\cite{lbs} targets the SDR middle to low bitrate range video contents. The SDR datasets, AVT-VQDB-UHD-1 \cite{avt} and LBS, are used as an independent cross-domain tests in Section \ref{sec:cross_dataset}. 
Each of these databases was created using the absolute category rating with hidden reference (ACR-HR) method\cite{p910}. To derive clean Mean Opinion Scores (MOS) from the raw ratings, we employed the P.910 method\cite{sureal}.

\begin{figure}[!htb]
\centering
\begin{subfigure}{.23\textwidth}
  \centering
  \includegraphics[width=\linewidth]{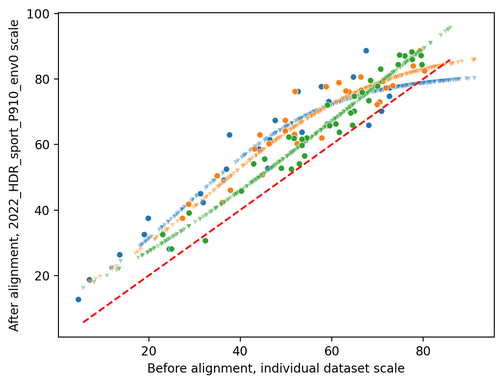}
\end{subfigure}
\begin{subfigure}{.23\textwidth}
  \centering
  \includegraphics[width=\linewidth]{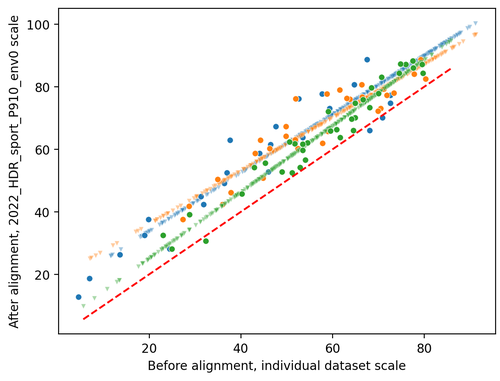}
\end{subfigure}
\caption{Non-linear (left) and linear (right) combination of subjective datasets using anchor videos. The dots are the common anchor videos between two datasets. Blue is HDR-SDR, Orange is HDR-LIVE, Green is HDR-AQ. All datasets are mapped to HDR-Sports scale. Red line is identical mapping.}
\label{fig:combine}
\end{figure}

To build a large scale databases, we combined the 4 HDR datasets linearly using common anchor videos which are the same set of videos rated during the creation of each respective datasets. We tested three  approaches: direct (identical), non-linear\cite{hdr_sports} and linear\cite{TransmissionRating, pc_to_uqs, combineds} combinations. As shown in Figure \ref{fig:combine}, the non-linear combination exhibits a saturation effect due to the inadequate coverage of the quality range by the anchor videos. E.g, the mapped HDR-SDR video scores (blue dots) are below 80. In contrast, the linear combination can map studies from varied environments to a consistent scale. Our VQM training result also shows linear mapping is the best. Therefore we used the linearly combined data as our training set. The resulting combined HDR dataset have 1614 videos on the unified HDR-Sports in-lab environment scale.

\section{Encoder-Quantization-Motion Quality Metrics}
The Encoder-Quantization-Motion Video Quality Metrics (EQM) is a lightweight encoder-generated objective video quality metric that computed from a set of quantization and motion related encoding features of the High Efficiency Video Coding (HEVC) standard. 
Similar to P.1204 standard\cite{p1204}, we benchmarked EQM on three levels to give different compute-accuracy trade-off:
\begin{enumerate}
    \item Metadata Level: It uses partial video-level metadata features from EQM. It is the fastest but less precise.
    \item No Reference Level (NR): This is the primary focus of this work. We introduced a new set of HEVC features derived from encoding or decoding processes. There's no need for any pixel-domain processing to determine the quality scores at this level, making it ideal for real-time latency-sensitive scenarios.
    \item Full Reference Level (FR): This level combines EQM features with pixel-based detail loss metric (DLM) \cite{dlm} feature. Since it offers the highest correlation, it's well suited for off-line quality optimization application.
\end{enumerate}

\subsection{Metadata Feature Set}
\label{sec:meta}
The metadata features consist of video-level attributes that can be quickly extracted from the media header, eliminating the need for full decoding. These features provide a fast baseline for objective quality assessment. As detailed in Table \ref{tab:metadata}, these features encompass common video properties.

\begin{table}[] 
    \centering
    \scalebox{0.7}{
    \begin{tabular}{|l|l|}
    \hline
    Feature Name       & Descriptions                                                                                                                                                            \\ \hline
    Resolution         & Resolution in terms of pixel count in a video                                                                                                                           \\ \hline
    FrameRate          & Frames per second                                                                                                                                                       \\ \hline
    Codec              & Video codec, e,g, h264, h265                                                                                                                                            \\ \hline
    PixelFormat      & \begin{tabular}[c]{@{}l@{}}Including chroma subsampling method, bitdepth, \\ and pixel range (limited/full), e.g. yuv420p10le, yuv420p\end{tabular}                     \\ \hline
    Bitrate            & Average bitrate of the video segment                                                                                                                                    \\ \hline
    \end{tabular}
    }
    \caption{Video-level metadata features}
    \label{tab:metadata}
\end{table}

\subsection{Encoding Quantization Motion Feature Set}
\label{sec:bitstream_features_set}
We derived frame-level features based on the HEVC encoding processes. These features were then aggregated over time using various statistical methods (mean, min, max, kurtosis, etc.) to produce segment-level features. These segment-level features serve as the input for a machine learning (ML) model, determining a segment-level video quality score. For an overall video score, the features are averaged across all segments and input into the same ML model to yield a final score.

Table \ref{tab:bitstream_features} details our encoding feature pool. We utilized Quantization Parameter (QP) at minimum coding block (CB) size  granularity to identify distortions due to quantization and Motion Vector (MV) length-related features at minimum Prediction Unit (PU) size granularity to represent the impacts of motion magnitude on video quality. It's worth mention that encoder-side implementation could take the pre-quantized QP and MV for feature extraction while on decoder-side those information are not available. Block statistics features help identify spatial complexity and the quality masking effects of contrast and detail. MV angle features are incorporated to differentiate the block MV angle and the global MV angle, therefore measure quantization of the salient local random movement area. The Figure \ref{fig:features} shows the features selected in the model from the feature pool. 

\begin{table*}[]
\centering
\scalebox{0.65}{
\begin{tabular}{|c|c|l|c|c|}
\hline
Feature   Type                               & Feature Name       & \multicolumn{1}{c|}{Descriptions}                             & Block to Frame Pooling & Frame to Segment Pooling  \\ \hline
Frame   Size                                 & FrameSize          & The final encoded frame size for I/P/B   frames               & N/A                    & mean/std/kurtosis/min/max \\ \hline
\multirow{3}{*}{QP}                          & minQP            & Minimum QP of all blocks in a frame                           & min                    & 
Interquartile range (IQR)                       \\ \cline{2-5} 
                                             & maxQP            & Maximum QP of all blocks in a frame                           & max                    & std                       \\ \cline{2-5} 
                                             & AvgQP             & Mean QP of all blocks in a frame                              & mean                   & mean/std/kurtosis/min/max \\ \hline
\multirow{2}{*}{Block Statistics}            & AvgBlockDepth       & The area-weighted average block depth                         & mean                   & median/kurtosis           \\ \cline{2-5} 
                                             & SkipBlksRatio      & The percentage of skipped blocks in B and P frames            & percentage             & median/kurtosis           \\ \hline 
\multirow{3}{*}{MV Length}                   & StdDevMotion     & The standard deviation of motion vector length for each frame & standard deviation     & mean                      \\ \cline{2-5} 
                                             & AvgMotion         & The average motion vector length for each frame               & mean                   & mean/kurtosis             \\ \cline{2-5} 
                                             & AvgQpLm             & Average QP of low motion blocks                                  & mean                   & std                    \\ \hline
MV Angle                                     & AvgQpLocalMvDir    & Average QP for local random motion blocks                     & weighted average       & mean                      \\ \hline
\end{tabular}
}
\caption{Segment-level encoder-based features designed for EQM NR mode}
\label{tab:bitstream_features}
\end{table*}

\subsubsection{Motion Length Features}
The average motion vector (MV) length serves as an indicator of motion magnitude. During the encoding motion estimation process, the encoder chooses the optimal MV based on the trade-off between rate and distortion. By extracting this information from the encoding or decoding stages, we can capture the impact of motion on video quality without using the costly optical flow estimation.

It's crucial to normalize MV length to account for variables like distance to the reference frame, resolution, and frame rate. This normalization ensures MV lengths can be compared consistently across diverse videos.
Given that the MV is expressed in units of a 1/4 pixel for luma, higher resolution videos require larger MVs to represent the same proportion of motion as videos with lower resolutions. Likewise, videos with higher frame rates have smaller MVs since each frame represents a shorter time span. %

\begin{equation}
F_{res} = \text{max\_frame\_width} / \text{frame\_width}
\end{equation}
\begin{equation}
F_{fr} = \text{frame\_rate} / \text{max\_frame\_rate}
\end{equation}

For P frames, $N_P$ is the normalization factor for different distance from the reference frame to the current frame.
\begin{equation}
N_{P} =  \frac{1}{|\text{POC}_c - \text{POC}_0|}
\end{equation}

\begin{equation}
\text{AvgMotion}_P = F_{fr} F_{res} \sum_{i}(N_{P}|\text{MV}|)
\end{equation}

Where $F_{fr}$ is frame rate factor, $F_{res}$ is resolution factor. $\text{POC}_c$ and $\text{POC}_0$ is the picture order count for the current frame and reference frame in list 0, $|\text{MV}|=\sqrt{\text{MV}_x^2 + \text{MV}_y^2}$ is the motion vector length of that 4x4 block within a PU. $i$ is for all the 4x4 blocks within the valid PU of a frame. 

For B frames, some blocks might have 2 motion vectors, therefore each of the MV is normalized base on their distance to the respective reference frame and then the two MV can be averaged.
The $\text{AvgMotion}_P$ and $\text{AvgMotion}_B$ are used as frame level MV length \texttt{AvgMotion} statistics, similar with the \texttt{StdDevMotion} for B and P frames. Then the frame level MV length features are pooled to segment level with mean or kurtosis as input to the VQM.

\subsubsection{Motion Angle features}
\begin{figure}
\centering
\includegraphics[width=0.7\linewidth]{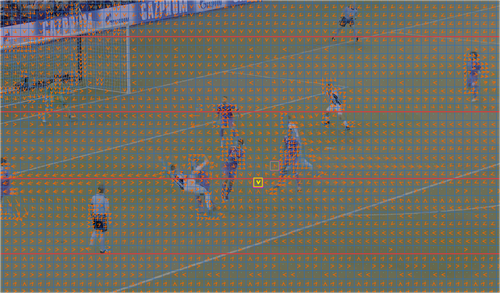}
\caption{Blocks with local random motion are more salient}
\label{fig:local_motion}
\end{figure}

The current MV feature only utilized the MV length but not the MV direction. The MV angle features can provide information about salient local motion. As shown in Figure \ref{fig:local_motion}, certain blocks exhibit random local motion, particularly noticeable in sports video clips around player regions. Such motions are distinct and are often critical for viewer experience. Our goal is to craft features that focus on MV direction, enabling the distinction of this vital local random motion from the overall global motion. When averaging QP, we could assign higher weights to those blocks dominated by this local randomness to better represent the saliency (fixation location) impacts on quality. It could be used in the 2nd pass in-loop saliency-aware encoding to improve video quality. 

\begin{figure}
\centering
\begin{subfigure}{.18\textwidth}
  \centering
  \includegraphics[width=\linewidth]{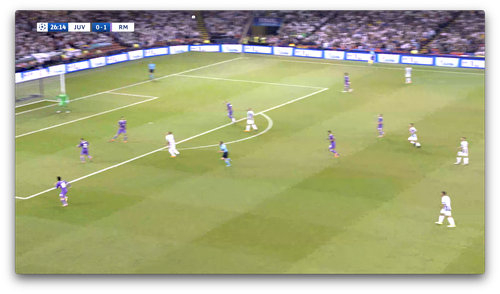}
  \caption{Video frame when camera panning left}
\end{subfigure}
\begin{subfigure}{.2\textwidth}
  \centering
  \includegraphics[width=\linewidth]{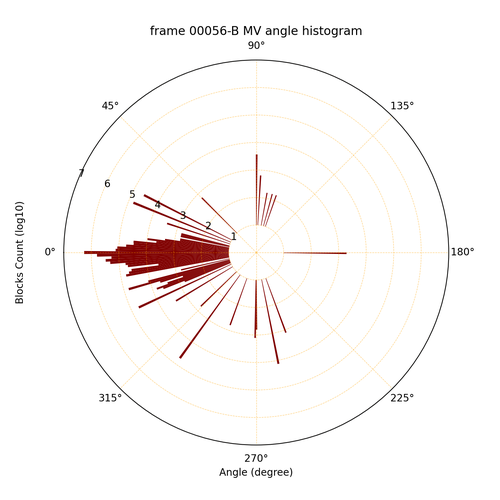}
  \caption{MV angles polar histogram in log10 scale}
\end{subfigure}
\caption{MV angle features on a B frame from a sports video.}
\label{fig:mv_features}
\end{figure}

We first calculate the direction of every PU blocks' MV in a B/P frame. Then we put the MV into 360 bins of each degree to have the MV block count polar histogram as shown in Figure \ref{fig:mv_features}. We can find that as the global camera motion panning left in the sport scene, the MV density is high on the left as well. 

Global Motion Direction:
Global motion direction are a set of angles (0 to 360) whose total block counts are higher than a threshold (80\%). While other directions are consider to be local motion directions. The feature \texttt{MvGlobalAngle} is defined as average of the 80\% global blocks’ angle. The average uses vector addition divided by the block count of that direction. Even though HEVC only model MV with simple translation with 2 degree of freedom, as shown in Figure \ref{fig:mv_features}, the global MV angle distribution can still be used for scene classification, e.g. pan and tile focus on single direction while zoom should show omnidirectional distribution. 

\begin{figure}
\centering
\includegraphics[width=0.85\linewidth]{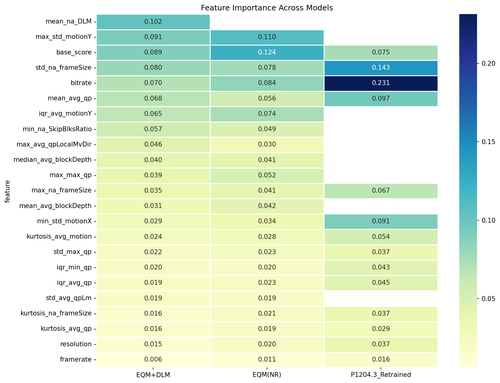}
\caption{Features and their importance in the RF residual model for EQM FR, EQM NR and P.1204.3 retrained model. The 1st statistics is frame-to-segment, the 2nd is block-to-frame.}
\label{fig:features}
\end{figure}

Average QP of local motion area:
The 20\% of local motion blocks are those whose MV angles are different from the top 80\% count global motion angles. In fact, after the features selection, the \texttt{AvgQpLocalMvDir} which usually represent the region of interest is being selected but not the less representative global MV direction counterpart. The feature \texttt{max\_avg\_qpLocalMvDir} also plays an important role for FR and NR model in Figure \ref{fig:features}. The random local motion blocks are usually high salient region in sports where the players are located as shown in Figure \ref{fig:local_motion}.

\subsubsection{Block Statistics Features}

\texttt{SkipBlksRatio} represents the percentage of skipped blocks (Zero MV difference, Zero transform coefficients) in a frame.
\texttt{AvgBlockDepth} is the area-weighted (with \(4 \times 4\) block granularity) average \(\log_2\) of the coding unit (CU) size, same as block depth. A smaller value indicates that the CU-Quad-Tree is split more finely, implying higher spatial-temporal complexity in terms of rate-distortion cost. %

\section{Model Performance}
We utilized a ML model similar to the P.1204.3\cite{p1204}, but with a fine-tuned Random Forest Regression (RF) model. EQM consists of two parts: the base model and the residual model. The base model takes metadata and averaged QP as input and outputs a quality score, which serves as input for the residual model. Differing from P.1204.3, we replaced the parametric model with a RF model and our model runs on the encoder-side with little compute overhead. The residual model incorporates additional motion, quantization and joint features for NR mode and pixel-based features for FR mode. 
As ablation study, we also evaluated performance when employing a single RF model without the residual learning. As demonstrated in Table \ref{tab:model_performance}, combining the residual model with the base model improved the model performance from row 7 to row 2, 5 to 3, and 4 to 1. This showcases how the residual learning approach substantially enhances model performance across various feature combinations.
We also add the feature importance heatmap in Fig. \ref{fig:features} for NR and FR level of EQM model and retrained P.1204.3. It can be shown that the base model output, bitrate, framesize, motion and average QP are important to the final quality prediction. For FR EQM the DLM\cite{dlm} features plays an especially important role. For NR EQM and P.1204.3 bitrate, frame size are more important.

\begin{table*}[]
    \centering
    \scalebox{0.8}{
    \begin{tabular}{|l|l|ll|l|l|l|}
    \hline
    ID & Model name                                  & \multicolumn{1}{l|}{Base model   features}               & Residual   features       & SROCC   & PLCC    & RMSE     \\ \hline
    \rowcolor[HTML]{FFFF00} 
    1  & EQM + DLM (FR) & \multicolumn{1}{l|}{\cellcolor[HTML]{FFFF00}metadata+QP} & EQM+DLM             & \textbf{0.96183} & \textbf{0.95987} & 5.80155  \\ \hline
    \rowcolor[HTML]{FFFF00} 
    2  & EQM (NR)       & \multicolumn{1}{l|}{\cellcolor[HTML]{FFFF00}metadata+QP} & EQM                 & 0.96162 & 0.95957 & 5.81790  \\ \hline
    3  & EQM + VMAF (FR) & \multicolumn{1}{l|}{metadata+QP}                         & EQM+VMAF            & 0.96158 & 0.95977 & \textbf{5.79331}  \\ \hline
    \rowcolor[HTML]{C6E0B4} 
    4  & RF EQM + DLM              & \multicolumn{1}{l|}{\cellcolor[HTML]{C6E0B4}N/A}         & EQM+DLM             & 0.95586 & 0.95291 & 6.39069  \\ \hline
    \rowcolor[HTML]{C6E0B4} 
    5  & RF EQM + VMAF              & \multicolumn{1}{l|}{\cellcolor[HTML]{C6E0B4}N/A}         & EQM+VMAF            & 0.95263 & 0.94991 & 6.49113  \\ \hline
    6  & P.1204.3 (Retrained)           & \multicolumn{1}{l|}{metadata+QP}                         & P.1204.3        & 0.95136 & 0.94937 & 6.49816  \\ \hline
    \rowcolor[HTML]{C6E0B4} 
    7  & RF EQM                    & \multicolumn{1}{l|}{\cellcolor[HTML]{C6E0B4}N/A}         & EQM                 & 0.95021 & 0.94662 & 6.79879  \\ \hline
    8  & EQM NR (w/o base QP)           & \multicolumn{1}{l|}{metadata}                            & EQM                 & 0.94904 & 0.94619 & 6.67085  \\ \hline
    \rowcolor[HTML]{C6E0B4} 
    9  & RF Metadata QP                                & \multicolumn{1}{l|}{\cellcolor[HTML]{C6E0B4}N/A}         & metadata+QP               & 0.94523 & 0.94209 & 6.95092  \\ \hline
    \rowcolor[HTML]{FFFF00} 
    10 & Metadata                                    & \multicolumn{1}{l|}{\cellcolor[HTML]{FFFF00}N/A}         & metadata                  & 0.90236 & 0.89656 & 9.12860  \\ \hline
    \rowcolor[HTML]{BDD7EE} 
    11 & VMAF                                        & \multicolumn{2}{c|}{\cellcolor[HTML]{BDD7EE}}                                        & 0.89310 & 0.87700 & 14.20221 \\ \cline{1-2} \cline{5-7} 
    \rowcolor[HTML]{BDD7EE} 
    12 & DLM                                        & \multicolumn{2}{c|}{\cellcolor[HTML]{BDD7EE}}                                        & 0.86101 & 0.83574 & 74.83098 \\ \cline{1-2} \cline{5-7} 
    \rowcolor[HTML]{BDD7EE} 
    13 & P.1204.3                      & \multicolumn{2}{c|}{\cellcolor[HTML]{BDD7EE}}                                        & 0.85689 & 0.85420 & 71.40934 \\ \cline{1-2} \cline{5-7} 
    \rowcolor[HTML]{BDD7EE} 
    14 & PSNR Y                                     & \multicolumn{2}{c|}{\multirow{-4}{*}{\cellcolor[HTML]{BDD7EE}Out-of-the-box}}        & 0.81748 & 0.67811 & 34.30432 \\ \hline
    \end{tabular}
    }
    \caption{Model Performance. The green rows are single RF model without residual learning. The yellow rows are the 3 levels EQM is benchmarked on. The blue rows are the pre-trained on SDR datasets or non-trainable VQM, others are all retrained on HDR Combined dataset. The metrics SROCC,PLCC,RMSE are average from 1000 times 5-fold cross validation. 
    The ``metadata'', ``EQM'' are the feature sets in section \ref{sec:meta} and \ref{sec:bitstream_features_set}, ``QP'' is the average QP feature, ``P.1204.3'', ``DLM'' and ``VMAF'' are features from P.1204.3 and VMAF open source implementation. 
    }
    \label{tab:model_performance}
\end{table*}

\subsection{Cross Dataset Test on SDR}
\label{sec:cross_dataset}
As observed in Table \ref{tab:model_performance}, retrained model always have higher performance than any pre-trained model due to the similar distribution of the test set as it is split from the training set. The high correlation might not be able to generalize well if the model is applied on new data. Therefore, we test the model performance when extrapolates to SDR. The Low-Bitrate Sports dataset \cite{lbs} is created to capture the impact of different frame rate on sports content, focusing on 720p and below quality range. The model is trained on the HDR combined dataset and test on the SDR HEVC part of the LBS dataset of 360 videos. From table \ref{tab:lbs_performance}, the EQM NR model shows highest SROCC. The FR model is not the best due to the training-testing domain shift from HDR to SDR. The RF model could overfit on the DLM features as it's more sensitive to the pixel value change from PQ to BT-709 Gamma while NR model using higher-level abstraction features like QP and MV statistics which are less sensitive to these changes.
On AVT-VQDB-UHD-1 \cite{avt} dataset in Table \ref{tab:avt_performance}, EQM shows similar performance as P.1204.3 even though it's trained on HDR.

\begin{table}[]
\centering
\scalebox{0.7}{
\begin{tabular}{|l|l|l|l|l|}
\hline
VQM                  & SROCC & PLCC  & KROCC & RMSE   \\ \hline
EQM NR              & \textbf{0.805} & 0.787 & \textbf{0.603} & 10.36  \\ \hline
EQM Metadata             & 0.801 & \textbf{0.795} & 0.600   & \textbf{10.187} \\ \hline
EQM FR DLM          & 0.801 & 0.788 & 0.600   & 10.347 \\ \hline
P.1204.3 (Retrained) & 0.765 & 0.743 & 0.561 & 11.247 \\ \hline
P.1204.3             & 0.720 & 0.702 & 0.522 & 11.958  \\ \hline
DLM                  & 0.506 & 0.581 & 0.388 & 13.671 \\ \hline
VMAF                 & 0.454 & 0.522 & 0.341 & 14.326 \\ \hline
PSNR Y               & 0.126 & 0.303 & 0.088 & 16.013 \\ \hline
\end{tabular}
}
\caption{Cross dataset testing on Low-Bitrate Sports dataset 360 videos. Metrics is calculated against P.910 DMOS. EQM and retrained Model is trained on HDR and tested on SDR dataset.}
\label{tab:lbs_performance}
\end{table}

\begin{table}[]
\centering
\scalebox{0.7}{
\begin{tabular}{|c|c|c|c|c|}
\hline
VQM         & SROCC & PLCC  & KROCC & RMSE   \\ \hline
P.1204.3    & \textbf{0.982} & \textbf{0.987} & \textbf{0.912} & \textbf{3.592}  \\ \hline
EQM         & 0.980 & 0.986 & 0.895 & 3.718  \\ \hline
DLM         & 0.918 & 0.924 & 0.792 & 8.674  \\ \hline
VMAF 4K NEG & 0.897 & 0.917 & 0.766 & 9.043  \\ \hline
VMAF        & 0.846 & 0.819 & 0.712 & 12.99  \\ \hline
PSNRY       & 0.846 & 0.794 & 0.710 & 13.877 \\ \hline
\end{tabular}
}
\caption{Cross dataset testing on AVT-VQDB-UHD-1 test 1 open-source 40 videos. EQM is trained on HDR while others are trained on SDR and tested on SDR dataset.}
\label{tab:avt_performance}
\end{table}

\subsection{Computation Efficiency}
On the encoder side, many features required to calcualte the EQM can be accumulated at the block-level and calculated at the end of the frame with minimum overhead. We tested our encoder integration on a 8-second 4K HDR10 source video from HDR-AQ dataset on single-thread x265 medium preset on AWS c7a.4x EC2. As shown in Table \ref{tab:encoder_runtime}, the calculation overhead at encoder is around 2\%-6\% of the encoding time. 
\begin{table}[t]
\centering
\scalebox{0.7}{
\begin{tabular}{|c|c|c|c|c|}
\hline
Resolution &
  \begin{tabular}[c]{@{}c@{}}Bitrate \\ (kbps)\end{tabular} &
  \begin{tabular}[c]{@{}c@{}}x265 w/o \\ EQM (s)\end{tabular} &
  \begin{tabular}[c]{@{}c@{}}x265 w \\ EQM (s)\end{tabular} &
  Overhead \%  \\ \hline
3840x2160 & 25000 & 368.9 & 376.0 & 1.9\% \\ \hline
3840x2160 & 3000  & 149.5 & 157.8 & 5.2\% \\ \hline
1920x1080 & 6000  & 69.4  & 71.6  & 3.1\% \\ \hline
1920x1080 & 1000  & 34.8  & 36.9  & 5.9\% \\ \hline
\end{tabular}
}
\caption{EQM encoder calculation runtime performance on a test video with different resolutions and bitrates}
\label{tab:encoder_runtime}
\end{table}

On the decoder-side, we implemented the frame-level multi-threading in our video parser built on top of ffmpeg HEVC decoder. The EQM research prototype implementation can achieve on average 53.4 fps compared to 19.97 fps from VMAF on 360 HEVC videos from LBS datasest using 15 threads on AWS c6a.4x EC2.

\section{Future Works}
This model only capture the downscaling and compression quality degradation of HEVC but can be extended to other codecs. Further improvement can be made from the encoder side. E.g. QP only measures the quantization step size. Sum of absolute difference (SAD) can be used to measure both prediction error and quantization error. MV features can be replaced with more accurate motion vector from hierarchical motion estimation without considering coding rate cost.

\section{Summary}
By linearly combining 4 video quality datasets using anchor videos, we are able to build a large-scale heterogeneous dataset. Targeting the HEVC compression and scaling degradation, the proposed EQM can run on encoder with minimal compute overhead and give better correlation and crossover accuracy. The proposed feature set can be efficiently extracted from server/encoder side or client/decoder side to boost performance of other VQMs. Moreover, those features can be better extrapolated to different dynamic range compared to the pixel-based features. 

{
\small
\bibliographystyle{IEEEtran}
\bibliography{reference}

\begin{thebibliography}{10}
\providecommand{\url}[1]{#1}
\csname url@samestyle\endcsname
\providecommand{\newblock}{\relax}
\providecommand{\bibinfo}[2]{#2}
\providecommand{\BIBentrySTDinterwordspacing}{\spaceskip=0pt\relax}
\providecommand{\BIBentryALTinterwordstretchfactor}{4}
\providecommand{\BIBentryALTinterwordspacing}{\spaceskip=\fontdimen2\font plus
\BIBentryALTinterwordstretchfactor\fontdimen3\font minus \fontdimen4\font\relax}
\providecommand{\BIBforeignlanguage}[2]{{%
\expandafter\ifx\csname l@#1\endcsname\relax
\typeout{** WARNING: IEEEtran.bst: No hyphenation pattern has been}%
\typeout{** loaded for the language `#1'. Using the pattern for}%
\typeout{** the default language instead.}%
\else
\language=\csname l@#1\endcsname
\fi
#2}}
\providecommand{\BIBdecl}{\relax}
\BIBdecl

\bibitem{SSIM}
Z.~Wang, A.~C. Bovik, H.~R. Sheikh, and E.~P. Simoncelli, ``Image quality assessment: from error visibility to structural similarity,'' \emph{IEEE transactions on image processing}, vol.~13, no.~4, pp. 600--612, 2004.

\bibitem{MSSIM}
Z.~Wang, E.~Simoncelli, and A.~Bovik, ``Multiscale structural similarity for image quality assessment,'' in \emph{The Thrity-Seventh Asilomar Conference on Signals, Systems \& Computers, 2003}, vol.~2, 2003, pp. 1398--1402 Vol.2.

\bibitem{NIQE}
A.~Mittal, R.~Soundararajan, and A.~C. Bovik, ``Making a “completely blind” image quality analyzer,'' \emph{IEEE Signal Processing Letters}, vol.~20, no.~3, pp. 209--212, 2013.

\bibitem{li2016vmaf}
Z.~Li, A.~Aaron, I.~Katsavounidis, A.~Moorthy, and M.~Manohara, ``Toward a practical perceptual video quality metric,'' \emph{The Netflix Tech Blog}, vol.~6, no.~2, 2016.

\bibitem{p1204}
A.~Raake, S.~Borer, S.~M. Satti, J.~Gustafsson, R.~R.~R. Rao, S.~Medagli, P.~List, S.~Göring, D.~Lindero, W.~Robitza, G.~Heikkilä, S.~Broom, C.~Schmidmer, B.~Feiten, U.~Wüstenhagen, T.~Wittmann, M.~Obermann, and R.~Bitto, ``Multi-model standard for bitstream-, pixel-based and hybrid video quality assessment of uhd/4k: Itu-t p.1204,'' \emph{IEEE Access}, vol.~8, pp. 193\,020--193\,049, 2020.

\bibitem{hdr_live}
Z.~Shang, J.~P. Ebenezer, A.~C. Bovik, Y.~Wu, H.~Wei, and S.~Sethuraman, ``Subjective assessment of high dynamic range videos under different ambient conditions,'' 2022.

\bibitem{hdr_aq}
Z.~Shang, J.~Ebenezer, Y.~Chen, Y.~Wu, H.~Wei, S.~Sethuraman, and A.~Bovik, ``A subjective and objective study of adaptive quantization of hdr videos,'' \emph{Submitted to IEEE Transactions on Circuits and Systems for Video Technology}, 2024.

\bibitem{hdr_sports}
Z.~Shang, Y.~Chen, Y.~Wu, H.~Wei, and S.~Sethuraman, ``Subjective and objective video quality assessment of high dynamic range sports content,'' in \emph{Proceedings of the IEEE/CVF Winter Conference on Applications of Computer Vision (WACV) Workshops}, January 2023, pp. 556--564.

\bibitem{hdr_sdr}
J.~P. Ebenezer, Z.~Shang, Y.~Chen, Y.~Wu, H.~Wei, S.~Sethuraman, and A.~C. Bovik, ``Hdr or sdr? a subjective and objective study of scaled and compressed videos,'' 2023.

\bibitem{lbs}
J.~P. Ebenezer, Y.~Chen, Y.~Wu, H.~Wei, and S.~Sethuraman, ``Subjective and objective quality assessment of high-motion sports videos at low-bitrates,'' in \emph{2022 IEEE International Conference on Image Processing (ICIP)}, 2022, pp. 521--525.

\bibitem{avt}
R.~R. Ramachandra~Rao, S.~Göring, W.~Robitza, B.~Feiten, and A.~Raake, ``Avt-vqdb-uhd-1: A large scale video quality database for uhd-1,'' in \emph{2019 IEEE International Symposium on Multimedia (ISM)}, 2019, pp. 17--177.

\bibitem{p910}
T.~Installations and L.~Line, ``Itu-tp. 910,'' \emph{Subjective video quality assessment methods for multimedia applications, Recommendation ITU-T}, 2023.

\bibitem{sureal}
\BIBentryALTinterwordspacing
Z.~Li, C.~G. Bampis, L.~Krasula, L.~Janowski, and I.~Katsavounidis, ``A simple model for subject behavior in subjective experiments,'' 2020. [Online]. Available: \url{https://arxiv.org/abs/2004.02067}
\BIBentrySTDinterwordspacing

\bibitem{TransmissionRating}
P.~Pérez, ``The transmission rating scale and its relation to subjective scores,'' in \emph{2023 15th International Conference on Quality of Multimedia Experience (QoMEX)}, 2023, pp. 31--36.

\bibitem{pc_to_uqs}
M.~Pérez-Ortiz, A.~Mikhailiuk, E.~Zerman, V.~Hulusic, G.~Valenzise, and R.~K. Mantiuk, ``From pairwise comparisons and rating to a unified quality scale,'' \emph{IEEE Transactions on Image Processing}, vol.~29, pp. 1139--1151, 2020.

\bibitem{combineds}
\BIBentryALTinterwordspacing
Y.~Pitrey, U.~Engelke, M.~Barkowsky, R.~P{\'e}pion, and P.~Le~Callet, ``{Aligning subjective tests using a low cost common set},'' in \emph{{Euro ITV}}, Lisbonne, Portugal, Jun. 2011, p. irccyn contribution. [Online]. Available: \url{https://hal.science/hal-00608310}
\BIBentrySTDinterwordspacing

\bibitem{dlm}
S.~Li, F.~Zhang, L.~Ma, and K.~N. Ngan, ``Image quality assessment by separately evaluating detail losses and additive impairments,'' \emph{IEEE Transactions on Multimedia}, vol.~13, no.~5, pp. 935--949, 2011.

\end{thebibliography}
}

\end{document}